\documentclass[twoside]{bour}
\usepackage{ritort_h}

\usepackage{epsfig}

\newcommand{\bi}{\begin{itemize}}
\newcommand{\ei}{\end{itemize}}
\newcommand{\be}{\begin{equation}}
\newcommand{\ee}{\end{equation}}

\newcommand{\bea}{\begin{eqnarray}}
\newcommand{\eea}{\end{eqnarray}}
\newcommand{\beastar}{\begin{eqnarray*}}
\newcommand{\eeastar}{\end{eqnarray*}}

\newcommand{\eq}[1]{~(\ref{#1})}
\newcommand{\eqq}[2]{~(\ref{#1},\ref{#2})}
\newcommand{\eqqq}[3]{~(\ref{#1},\ref{#2},\ref{#3})}

\newcommand{\kuf}{k_{u\to f}}

\newcommand{\kfu}{k_{f\to u}}
\newcommand{\xfu}{\Delta x_{f\to u}}
\newcommand{\xuf}{\Delta x_{u\to f}}
\newcommand{\dF}{\Delta F_0}

\begin{document}

\title{Work fluctuations, transient violations of the second law and
free-energy recovery methods:\\ Perspectives in Theory and Experiments}

\author{F\'elix {\sc Ritort}\\   Departament of Physics\\
 Faculty of Physics\\ University of Barcelona,\\ Diagonal 647\\ 08028
Barcelona, Spain \\
{\it and}\\
Department of Physics\\ University of California\\ Berkeley CA 94720, USA}

\maketitle

\begin{abstract} In this report I discuss fluctuation theorems and
transient violations of the second law of thermodynamics in small
systems. Special emphasis is placed on free-energy recovery methods in
the framework of non-equilibrium single molecule pulling experiments. The
treatment is done from a unified theoretical-experimental perspective
and emphasizes how these experiments contribute to our understanding
of the thermodynamic behavior of small systems.
\end{abstract}


\section{Biophysics and statistical physics}
Living systems are the most notable example of how matter can organize
into states of extremely high complexity. The investigation of the
structural organization of biological matter was boosted since the
discoveries of the double helix structure of the DNA by Watson and
Crick and the ensuing discovery of the structure of various proteins
half century ago. Microscopic and spectroscopic techniques have
greatly developed since then and current research is revealing an
unprecedented richness of details about the functional behavior of
living systems at the molecular and cellular level.

Biophysics is an area of science at the interface of physics,
chemistry and biology. It is probably the most important
interdisciplinary area of research whose knowledge requires a good
understanding of how matter functions at the physical, chemical or
biological level. While physics during the past has traditionally
avoided the study of complex systems as imperfect, unapproachable or
even uninteresting, the fact is that complexity is becoming a more and
more common abode for physicists ~\cite{LauPinSchStoWol00}.  Among all
possible disciplines in physics, statistical physics occupies a
privileged position as the natural framework to understand the
behavior of biological systems at the molecular level. Stochasticity,
fluctuations, metastability and thermal activation are concepts that
are commonly used in statistical physics, yet they are also relevant
to understanding the great variety of tasks carried out by
biomolecules.

Thermodynamics is the discipline that describes the
exchange processes of energy and matter that occur at the molecular and
cellular level. However, thermodynamics, a science inherited in the 18th
century from the times of the industrial revolution, has been inspired
by motors and steam engines that proved to be indispensable during that
time. It is fair then to question the relevance and applicability of all
this knowledge when scientists immerse into the realm of the very small,
far from the initial context that inspired Carnot and others.
There has been a recent interest in the study of the so called work
fluctuations and transient violations of the second law in systems
driven to a non-equilibrium state. Fluctuation theorems quantify the
probability of those non-equilibrium trajectories that, taken
individually, violate some of the inequalities of thermodynamics.  For
macroscopic systems these trajectories are known to be irrelevant and
unobservable, however at the level of the small, when the energies
interested are of order of several times $k_BT$, these rare
trajectories might become important. Although thermodynamic
inequalities are known to describe the behavior of average values, it
is important to explore the implications and relevance of these
deviations in our understanding of energy transformation processes at
the molecular level.  A quantitative experimental observation and
measurement of these trajectories has only recently become
possible. This report describes these experiments from a unified theoretical-experimental
perspective and emphasizes how these experiments contribute to our
understanding of the thermodynamic behavior of small systems.

Sec.~\ref{thermo} is a short reminder about the second law of
thermodynamics. It serves to explain the importance of fluctuations in
small systems and short times. Section~\ref{fluct} describes work
fluctuations in the framework of stochastic systems. Particular
emphasis is put in the case where the system, initially in an
equilibrium state, is perturbed arbitrarily far from equilibrium. We
then discuss the non-equilibrium work relation originally derived by
Jarzynski. Sec.~\ref{expwork} describes the current state of the art
regarding experimental measurements of work
fluctuations. Sec.~\ref{sm} presents a digression on single molecule
experiments as an excellent framework to investigate work fluctuations
and free-energy recovery methods applied to
biomolecules. Sec.~\ref{RNAexp} describes some of the experiments
conducted in the unfolding of small RNA molecules under the action of
an external force and the test of the Jarzynski
equality. Sec.~\ref{modexp} illustrates a model where work
fluctuations can be analytically computed as well as a comparison with
the experiments reported in the preceding section. Finally,
Sec.~\ref{conc} presents some conclusions and perspectives.

\section{Few facts about the second law of thermodynamics}
\label{thermo}
\begin{figure}
\begin{center}
\epsfig{file=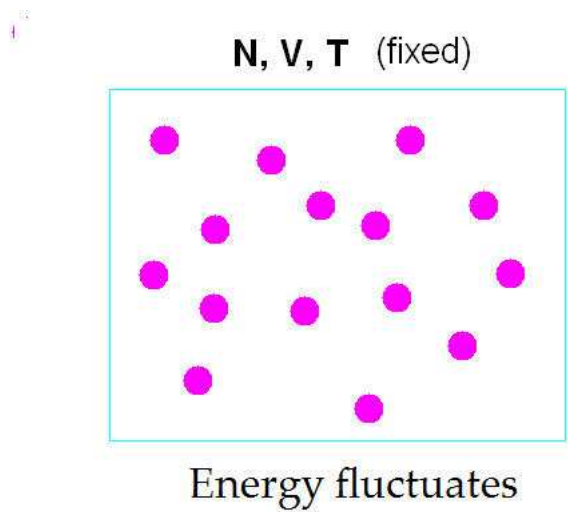,angle=0, scale=0.45}
\end{center}
\caption{Schematic representation of a vessel containing molecules 
that exchange energy and momentum with the surrounding bath through different
mechanisms such as collisions with the walls. If the number of
particles is small pressure fluctuations (due to the fluctuations in
the collision rate of the molecules against the walls) could be
observable with highly sensitive instruments.}
\label{fig.gas}
\end{figure}

To put in perspective the content of the present article we start by
recalling few facts about the second law of thermodynamics
\cite{Callen85,Atkins84}.  Let us consider a gas consisting of $N$
molecules enclosed in a given vessel of volume $V$. The vessel is in contact with a thermal bath at
temperature $T$ and the gas inside is kept in equilibrium (i.e. its
macroscopic properties remain stationary), see Fig.~\ref{fig.gas}.
Particles in the gas collide with the walls of the container exerting
a pressure $P$ that is function of the volume $V$ and the temperature
$T$. Their relation defines the equation of state of the gas. In these
conditions heat is continuously exchanged between the gas and the bath
through the walls of the vessel.  The state of the gas can be modified
by changing (e.g. expanding) the volume of the container from an initial
volume $V_i$ to a final volume $V_f$. If the transformation is done by
keeping the system always in contact with the bath at temperature $T$
the process is called isothermal. If the transformation is done slow
enough then the gas goes through a sequence of equilibrium states and
the process is called reversible. In general the transformation will not
be reversible and the gas will be driven to a non-equilibrium state
after the volume has been expanded. For the transformation $V_i\to V_f$
the first law of thermodynamics states that energy is conserved,
\be
\Delta E=\Delta Q+W=\Delta Q+P\Delta V
\label{thermo1}
\ee
with $\Delta V=V_f-V_i$. From\eq{thermo1} we see that the variation of
the energy of the gas $\Delta E$ is the sum of the work exerted upon the
system $W$ plus the net heat supplied from the bath to the system
$\Delta Q$. The difference between heat and the other state variables
$E$ and $V$ is important. If the volume or energy characterize the thermodynamic
state of the system, the amount of heat contained does not as it is fully interchangeable with
work depending on the path followed during the transformation. In a
general transformation, part of the total work exerted upon the system
is lost and dissipated in the form of heat to the surroundings. This is the
content of the second law as stated by Clausius,
\be
\Delta Q\le T\Delta S
\label{thermo2}
\ee
where $S(V,T)$ is a state function called entropy.  The amount of heat
lost during the process is called dissipated work $W_{\rm dis}$. It is
given by the difference between the maximum amount of heat that can be
supplied to the system and the actual heat supplied (i.e. the right and left hand
sides of\eq{thermo2}),
\be
W_{\rm dis}=T\Delta S-\Delta Q~~~~~{\rm with}~~~~ W_{\rm dis}\ge 0~~~~.
\label{thermo3}
\ee
Another way to state the content of the second law is in terms of the
Helmholtz free energy $F=E-TS$. Using\eqqq{thermo1}{thermo2}{thermo3} we
have,
\be
\Delta F=\Delta E-T\Delta S=\Delta Q+W-T\Delta S=W-W_{\rm dis}=W_{\rm rev}
\label{thermo4}
\ee
where $W_{\rm rev}$ is the so-called reversible work, identical to the
free energy change $\Delta F$ associated to the initial and final
equilibrium states.  Only in a reversible transformation the
equality\eq{thermo2} is satisfied and $W=W_{\rm rev}$ or $W_{\rm
dis}=0$.

It has been known since the early days of statistical mechanics that
the work can fluctuate~\footnote{The existence of rare fluctuations
have given rise to several paradoxes, see~\cite{history} for an
historical perspective}. To better clarify what we mean by this let us
go back to the previous example and consider the gas of molecules
enclosed in the vessel depicted in Fig.~\ref{fig.gas}. Let us imagine
that we repeat many times the experiment of the expansion of the
container from $V_i$ to $V_f$ by following always the same variation
protocol $V(t)$ where $t$ is the time and the whole expansion lasts
for a time $t_0$~\footnote{The simplest protocol would be an expansion
of the volume at a constant rate $r$, $\dot{V}=r=\frac{\Delta V}{t_0}$
or $V(t)=V_i+rt$}. Then for each experiment a different work value
$W=\int PdV$ would be obtained as the pressure itself is a fluctuating
variable.  For a macroscopic system pressure fluctuations are
unobservable due to the large number of molecules (of the order of the
Avogadro number $N_A\sim 10^{23}$). However, for small systems,
fluctuations could be observable. The result that intensive quantities
display thermally induced fluctuations was put forward by
Niquist for the case of voltage fluctuations across a
resistance~\cite{Nyquist28} and generalized later on by Welton and
Callen~\cite{WelCal51}. Let us focus on the case where the gas
contains about 100 molecules and is kept in equilibrium inside a
volume $V$ at the room temperature $T=298K$.  Let us imagine that at
every interval of time (e.g. $\tau=1$ millisecond) we count the number
of collisions of the molecules against the wall, $N_c(\tau)$, as well
as the average momentum $<p_c>$ transferred by the colliding molecules
to the walls along each time interval. Then, the pressure exerted by
the molecules on the walls $P$ will be proportional to $N_c(\tau)$ and
$<p_c>$. An histogram of the values of $P$ thereby collected would
show a Gaussian distribution centered around a mean value $P_{\rm
mean}$ (which would also coincide with the most probable value) and a
variance that decreases like $1/N$. This is the content of the law of
large numbers. Sometimes the measured pressure will be large,
sometimes it will be small: the fluctuations will become observable as
the number of molecules is reduced. Initially these fluctuations will
show a characteristic Gaussian profile. However, upon reduction of the
number of molecules, deviations from the Gaussian behavior will be
observed. The same happens in the non-equilibrium experiment where the
volume is expanded. If the time $t_0$ is short enough or the number of
molecules $N$ is small enough, the work $W=\int PdV$ exerted upon the
system will fluctuate from one non-equilibrium trajectory to another,
and the fluctuations will become more noticeable as $t_0$ and $N$
decrease. If the gas is initially in an equilibrium state,
then\eqq{thermo2}{thermo3} holds in average. However, for some
trajectories the work will be such that $W_{\rm dis}<0$ and the
inequality\eqq{thermo2}{thermo3} will be reversed. These particular
set of trajectories are called violating trajectories or transient
violations of the second law.

The theory describing thermal fluctuations in equilibrium systems was
put forward long ago by Einstein. However, this theory describes
fluctuations of extensive quantities (such as the energy) in different
ensembles in the macroscopic regime where fluctuations are subleading
of order ${\cal O}(\sqrt{V})$, yet large. The relative magnitude of
these fluctuations compared to the actual value of the energy content
is of order ${\cal O}(1/\sqrt{V})$. In macroscopic samples relative
fluctuations are of order ${\cal O}(1/\sqrt{N_A})$. This gives
estimates for the magnitude of relative fluctuations of the order of
$10^{-11}$.  A theory of fluctuations for small systems (such as those
relevant to biophysics) where relative fluctuations are much larger
(e.g. of order one) should be explored if the theory of fluctuations,
as has been developed for macroscopic systems, reveals inadequate to
explain the behavior observed in future experiments.

\section{Fluctuation theorems in non-equilibrium systems}
\label{fluct}
As we discussed in the previous section the work is quantity that
fluctuates among different repetitions of the same experiment. Moreover,
as compared to other quantities such as the internal energy, the entropy or the
heat transferred, the amount of work exerted upon the system is a directly measurable
quantity. A relevant question is then to ask what
can we learn by measuring work fluctuations.

To answer this question we consider the dynamical evolution of a
system in contact with a large bath or reservoir where heat and/or
matter can be continuously exchanged. Both system and bath may appear
inextricably linked and no partial description of the system can be
achieved without considering the behavior of the bath and its
interaction with the system. The mathematical
treatment of the combined system plus bath complex represents a formidable
theoretical challenge and some coarse-graining strategies must be
considered to tackle this question. A common strategy is to consider a
reduced level in the description of the dynamics of the system where
many details of the bath have been eliminated in favor of a few
number of parameters (such as its temperature, its pressure or its
chemical potential). These parameters are thought to be sufficient
to characterize the heat and/or matter exchange between system and
bath.  
After this reduction is adopted, few requirements have to be
imposed on the dynamics in order to reproduce many of the observable properties.
In particular, dynamics must be microscopically reversible (or satisfying
detailed balance) and ergodic (all configurations must be
accessible)~\footnote{Ergodicity is not an essential property if one
considers equilibrium restricted to a given region of phase space,
the sole condition is that all configurations contained in that region
of phase space must be accessible during the dynamical process.}
to guarantee that the system reaches thermal equilibrium after
long times so the net heat exchange between system and bath
asymptotically decays to zero.

\subsection{Work fluctuations in stochastic systems}
\label{wf}
It is in the framework of such coarse-grained dynamics that we want to
focus our discussion and investigate the origin of work fluctuations. A
prominent example of such reduced dynamics are stochastic Markov
processes. The content of this section will be useful to establish the
notation that we will use throughout the paper. We now deepen the
mathematical level of our discussion and consider a general system
described by an energy function $E({\cal C})$ where ${\cal C}$ is a
generic configuration (in the example in Sec.~\ref{thermo} of a gas of
$N$ molecules, ${\cal C}$ would stand for the positions and momenta of
all molecules inside the vessel) in contact with a bath at temperature
$T$. The dynamics are assumed to be discrete in time with elementary
step $\Delta t$. A trajectory of the system is characterized by the
sequence of configurations ${\cal T}\equiv \lbrace {\cal C}_k;0\le k\le
N_s \rbrace$ where $k$ is the index for the discrete time step and $N_s$
is the total number of time steps. The time corresponding to step $k$ is
then given by $t=k\Delta t$ with $t=0 (k=0)$ and $t_f (k=N_s/\Delta t)$
denoting the initial and final times respectively. The continuous-time
limit is recovered if $\Delta t\to 0, N_s\to\infty$ with $t_f$ finite.
Dynamics are then defined by the set of probabilities $P_k({\cal C})$ for
the system to be found at configuration ${\cal C}$ at time-step $k$. The
$P_k({\cal C})$ satisfy a master equation. For a Markov process the time
evolution of these probabilities depends upon the form of the rates
$W_k({\cal C'}|{\cal C})$, defined as the transition probability per
unit time to go from configuration ${\cal C}$ to ${\cal C'}$ at
time-step $k$. These rates are assumed to lead to an ergodic dynamics
(where any pair of configurations are always connected by at least one
trajectory) and satisfy the detailed balance condition,
\be
\frac{W_k({\cal C'}|{\cal C})}{W_k({\cal C}|{\cal C'})}=\exp\Bigl(-\beta(E({\cal C'})-E({\cal C}))\Bigr) 
\label{wf1}
\ee
where $\beta=1/k_BT$, $k_B$ being the Boltzmann constant. Under very
general conditions this dynamics guarantees that the system reaches a
stationary state where configurations are populated according to the
Boltzmann weight.  The solution to the master equation gives the time
evolution for the system. 

Now we will treat the case where the system is perturbed in a prescribed
way and consider the ensemble of all possible non-equilibrium
trajectories that start from an initial state characterized by the
distribution $P_0({\cal C})$.  Because dynamics is stochastic, it will
generate an ensemble of non-equilibrium trajectories by repeating the
same experiment many times~\footnote{\label{footmarkov}The same result
holds for deterministic (e.g. Hamiltonian) dynamics. In this case the
ensemble of non-equilibrium trajectories is determined by the ensemble
of initial configurations sampled with probability $P_0({\cal C})$. The
set of phase space points then behaves as an incompressible fluid, a
consequence of the Liouville theorem. Hamiltonian dynamics can be seen
as a particular limit of stochastic dynamics, where rates $W_k({\cal
C'}|{\cal C})$ vanish except along the constant energy surface $E({\cal
C})=E({\cal C'})$ and are deterministic, i.e. rates are different from
zero only for pairs of configurations ${\cal C},{\cal C'}$ connected by
the equations of motion. Dynamics is reversible and corresponds
to\eq{wf1} with $W_k({\cal C'}|{\cal C})=W_k({\cal C}|{\cal C'})$. The
case of Hamiltonian dynamics was originally addressed by Jarzynski in
his original derivation of the non-equilibrium work
relation~\cite{Jarzynski97a}. The stochastic case has been analyzed also
for general Markov processes by Crooks and
Jarzynski~\cite{Jarzynski97b,Crooks98a,Crooks98b} and for Langevin
dynamics by Kurchan~\cite{Kurchan98}. For a discussion of the
similarities and differences between deterministic and stochastic
dynamics see~\cite{Jarzynski02}.}. In addition to the configuration
${\cal C}$, and in order to characterize the perturbation protocol, we
introduce a parameter $\lambda$ that specifies the value of the control
parameter that is changed throughout the non-equilibrium
process~\footnote{For simplicity we only consider the case of one
control parameter. For many control parameters the generalization is
straightforward.}. An important remark is now in place.  The control
parameter is a variable that can change in time but does not fluctuate.
The temporal sequence of values $\lbrace\lambda_k;0\le k\le N_s \rbrace$
defines the perturbation protocol and this sequence of values never
changes from experiment to experiment. Somehow, the control parameter
plays a role akin to the temperature of the bath. In particular, for a
fixed value of $\lambda$, we require that dynamics is such that the
system asymptotically reaches the thermodynamic state corresponding to
that value of $\lambda$. To understand how rates depend on the value of
$\lambda$ we reason as follows. The control parameter
usually shifts the energy levels of the system according to the
relation,
\be E_{\lambda}({\cal C})=E({\cal C})-\lambda A({\cal C})
\label{wf2}
\ee 
where $A({\cal C})$ is the observable coupled to the parameter
 $\lambda$~\footnote{For instance, if $\lambda$ is a magnetic or
 gravitational field then $A$ stands for the magnetization and the
 height of the center of mass respectively}. The simplest assumption is then to enforce
detailed balance for the perturbed rates,
\be
\frac{W_{\lambda}({\cal C'}|{\cal C})}{W_{\lambda}({\cal C}|{\cal
C'})}=\exp\Bigl(-\beta(E_{\lambda}({\cal C'})-E_{\lambda}({\cal
C}))\Bigr)=\frac{W({\cal C'}|{\cal C})}{W({\cal C}|{\cal
C'})}\exp\bigl(-\beta\lambda\Delta A\bigr)
\label{wf3}
\ee
where we used \eq{wf1} in the r.h.s. and the definition $\Delta
A=A({\cal C'})-A({\cal C})$. We now consider the variation of energy along a given trajectory
$\Delta E({\cal T})=E_{\lambda_f}({\cal C}_f)-E_{\lambda_0}({\cal C}_0)$ where
${\cal C_0},{\cal C}_f$ are the initial and final configurations for
that trajectory and $\lambda_0,\lambda_f$ are the initial and final
values of the control parameter as defined by the protocol (trajectory
independent). From\eq{wf2} this is given by,
\be \Delta E({\cal T})=\Bigl[\sum_{k=0}^{N_s-1}(E_{\lambda_k}({\cal
C}_{k+1})-E_{\lambda_k}({\cal C}_k)\Bigr]
-\Bigl[\sum_{k=0}^{N_s-1}A({\cal C}_k)\Delta \lambda_k\Bigr]=\Delta
Q({\cal T})+W({\cal T})
\label{wf3b}
\ee
with $\Delta \lambda_k=\lambda_{k+1}-\lambda_k$. This decomposition
was proposed originally by Crooks~\cite{Crooks98a} to identify work and
heat by using the first law of thermodynamics\eq{thermo1}.
The first term in\eq{wf3b} is identified as the heat transferred from the bath
to the system and the second with the work exerted upon the system. We concentrate our attention on the
the work exerted upon the system along a given trajectory
${\cal T}$,
\be
W({\cal T})=\sum_{k=0}^{N_s-1}\Bigl(\frac{\partial  E_{\lambda}({\cal
C}_k)}{\partial \lambda}\Bigr)_{\lambda=\lambda_k}\Delta \lambda_k=-
\sum_{k=0}^{N_s-1}A({\cal C}_k)\Delta \lambda_k\equiv-\int_0^t ds \dot{\lambda}(s)A({\cal C}(s))ds
\label{wf4}
\ee
where we have
applied the continuous-time limit~\footnote{In the continuous-time limit, both $\lbrace {\cal
C}_k,\lambda_k\rbrace$ become ${\cal C}(t),\lambda(t)$ defining the real-time
trajectory and perturbation protocol respectively.} in the last term in the
r.h.s. of\eq{wf4}. As the trajectory is stochastic the work is a
fluctuating quantity that can be characterized by its probability
distribution ${\cal P}(W)$ defined as,
\be
{\cal P}(W)=\sum_{{\cal T}}P({\cal T})\delta(W-W({\cal T}))
\label{wf5}
\ee
where ${\cal T}$ stands for the trajectory and was already defined. The
importance of ${\cal P}(W)$ relies upon the fact that it is a quantity
that is experimentally measurable and therefore is suitable to
quantitatively characterize work fluctuations along non-equilibrium
trajectories.

\subsection{The fluctuation theorem (FT)}
\label{ft}
Fluctuation theorems (FTs) provide specific relations for the quantity
${\cal P}(W)$ in\eq{wf5} for general non-equilibrium processes.  In
fact, until now nothing was said about the type of non-equilibrium
process and the treatment given in the previous section was
general. We defined concepts such as the initial and final state, the
perturbation protocol $\lambda(t)$, the trajectory ${\cal T}$ and the
work and heat along a given trajectory. The main difference between a
general non-equilibrium process and a reversible one is the enormous
and various type of situations one can encounter in the first
case. General physically meaningful statements about the properties of
the distribution\eq{wf5} quite probably do not exist and a specific
type of non-equilibrium process has to be adopted to come up with
specific results. Several fluctuation theorems have appeared in the
literature depending on the particular non-equilibrium context. Many
fall into the category of entropy production FTs. The first example in
this class was proposed by Evans, Cohen and Morriss~\cite{EvaCohMor93}
for systems in steady states. The entropy production there defined
bears some resemblance with the work\eq{wf4} that is exerted by the
external non-conservative forces that act upon the system. Several
related theoretical results have
followed~\cite{GalCoh95a,GalCoh95b,Maes99} as well as
experiments~\cite{CilLar98,Cil03}. A comprehensive review can be found
in \cite{EvaSea02}. Other more complex scenarios can be envisaged, for
example in the case where the system is in a non-stationary aging
state\footnote{Non-equilibrium aging states are widespread in
condensed matter physics. The most common example are structural
glasses quenched below their glass transition temperature. The aging
state is characterized by strong violations of the
fluctuation-dissipation theorem \cite{CriRit03a}.}.  In this case, no
work is performed upon the system and the relevant quantity turns out
to be the released heat from the system to the
bath~\cite{CriRit03b,Ritort03}.

The content of this article deals particularly with systems initially in
equilibrium with the bath that are driven to a non-equilibrium state by the action of an
applied perturbation~\cite{EvaSea94}. Therefore,
\be
P_0({\cal C})=P_{\rm eq}({\cal C})=\frac{\exp(-\beta E_{\lambda_0}({\cal
C})}{Z}
\label{ft1}
\ee
where $Z=\sum_{{\cal C}}\exp(-\beta E_{\lambda_0}({\cal C})$. This case
has been studied by Jarzynski \cite{Jarzynski97a,Jarzynski97b} and
Crooks \cite{Crooks98a,Crooks98b}. We omit details of the
derivation as these can be found in the references. A
particularly interesting identity has been derived by
Crooks~\cite{Crooks98b} who considered the forward and reverse paths in a
non-equilibrium process. The forward process ($F$) is characterized by
the protocol function $\lambda_F(t)$ with the initial state in
equilibrium at the value $\lambda_F(0)$. The reverse process ($R$) is
characterized by the inverted protocol function
$\lambda_R(t)=\lambda_F(t_0-t)$ with $t_0$ being the total time for the
forward process and the initial state for the reverse process being in
equilibrium at the value $\lambda_R(0)$ (which is equal to
$\lambda_F(t_0)$). The following result is obtained~\cite{Crooks98b},
\be
\frac{P_F(W)}{P_R(-W)}=\exp\Bigl(\frac{W-\Delta
F}{k_BT}\Bigr)=\exp\Bigl(\frac{W_{\rm dis}}{k_BT}\Bigr)~~~~.
\label{ft2}
\ee
Simple manipulation of this ratio and integration of one side of the
relation from $-\infty$ to $\infty$ gives,
\be
\int_{-\infty}^{\infty}P_F(W)\exp\Bigl(-\frac{W}{k_BT}\Bigr)=\exp(-\frac{\Delta
F}{k_BT}\Bigr)~~~~.
\label{ft3}
\ee 
This is the content of the Jarzynski equality originally derived in \cite{Jarzynski97a} for Hamiltonian
dynamics (see the footnote~\ref{footmarkov}).  In
what follows, if not stated otherwise, we will only consider the forward
process in a non-equilibrium experiment and drop the subscript $F$
for the work probability distribution $P_F$. We will use the symbol
$\overline{(...)}$ to denote the average over all non-equilibrium
trajectories generated by a given protocol. The Jarzynski equality (JE) reads,
\be
\overline{\exp\Bigl
(-\frac{W}{k_BT}\Bigr)}=\int_{-\infty}^{\infty}P(W)\exp\Bigl(-\frac{W}{k_BT}\Bigr)=
\exp(-\frac{\Delta F}{k_BT}\Bigr)~~~~.
\label{ft4}
\ee
\begin{figure}
\begin{center}
\epsfig{file=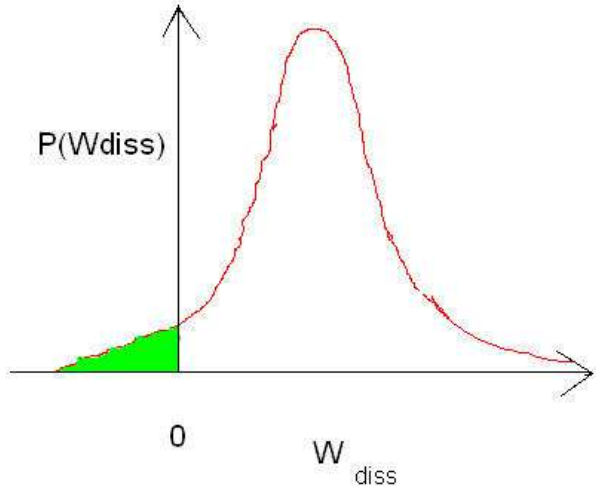,angle=0, scale=0.5}
\end{center}
\caption{A typical work probability distribution for a small system is
useful to characterize work fluctuations. Transient violations of the
second law are a particular class of trajectories with work values characterized
by the fact that the the Clausius inequality\eqq{thermo2}{thermo3} is reversed. 
}
\label{fig.pwfluct}
\end{figure}

This expression can be written in a more compact form,
\be
\overline{\exp\Bigl
(-\frac{W_{\rm dis}}{k_BT}\Bigr)}=1~~~~.
\label{ft5}
\ee
The Jarzynski equality provides a simple way to derive the second
law. Using Jensen's inequality \cite{Chandler87}  $\langle \exp(x)\rangle\ge \exp(\langle
x\rangle)$  in\eq{ft5} we obtain,
\be
\overline{W_{\rm dis}}\ge 0~~~~.
\label{ft6}
\ee
An important aspect of the JE~\eq{ft5} is that it introduces a
way to quantitatively estimate transient violations  of the second
law, i.e. the fraction of trajectories whose dissipated work is
negative. The reason is easy to understand by inspection of\eq{ft5}. The
average of the exponential in the l.h.s of \eq{ft5} equals 1 only if
trajectories with $\overline{W_{\rm diss}}<0$ exist. An analysis of the
JE in the near-equilibrium regime is useful. Close to equilibrium
the distribution $P(W)$ can be approximated by a
Gaussian~\footnote{\label{Gauss}The exact form of the $P(W)$ can be very
complicated, however the Gaussian approximation is expected to hold in
the non-equilibrium regime where work trajectories do not deviate much
from the reversible one. In this way the Gaussian approximation appears
tightly related to linear response theory. In the latter, quantities
deviate from their equilibrium values proportionally to the intensity of
the perturbation. In a similar way, the deviation of the average work
value (the first moment of $P(W)$) from its reversible value $\Delta F$
is expected to be linear with the perturbing speed $\dot{\lambda}$, see
also \cite{RitBusTin02} for a quantitative estimate of this statement. In other
special cases, however, the Gaussian result can be exact even for arbitrarily
strong perturbations. This is the example of a bead dragged through
water~\cite{MazJar99}.},
\be
P(W)=\frac{1}{\sqrt{2\pi \sigma^2}}\exp\Bigl( -\frac{(W-W_{\rm mean})^2}{2\sigma^2}\Bigr)
\label{ft7}
\ee
where $W_{\rm mean}$ is the value of the work at the center of the Gaussian and
$\sigma^2=\overline{W^2}-\overline{W}^2$ its variance.
Substitution of\eq{ft7} into\eq{ft4} gives,
\be
R=\frac{\sigma^2}{2\overline{W_{\rm dis}}k_BT}=1~~~~.
\label{ft8}
\ee
Throughout the paper we will refer to $R$ as the {\em
fluctuation-dissipation ratio} as it involves a ratio between work
fluctuations $\sigma_2$ and dissipation $\overline{W_{\rm
dis}}$~\footnote{The use of the term {\em fluctuation-dissipation ratio}
for the quantity $R$ has not to be confused with that adopted in glassy
systems and usually denoted by $x$~\cite{CriRit03a}. For glassy systems $x$
describes the ratio between a time derivative of the correlation
function and the response function. Albeit similar, the two-quantities
are not identical as they refer to different non-equilibrium
scenarios.}.  This result is a particular form of the
fluctuation-dissipation theorem in the linear response regime. In
general, $R\ne 1$ far from equilibrium. More about the dependence of $R$
with the value of the dissipated work will be said later in
Sec.~\ref{modexp}. We infer from\eq{ft8} that in the near-equilibrium
regime the variance of the work is of the same order of the average
dissipated work. In the reversible limit $W_{\rm dis}\to 0$ the JE
trivially holds as $W_{\rm dis}=0$ (or $W=\Delta F$).  In this limit the
number of trajectories with $W_{\rm dis}>0$ equals the number with
$W_{\rm dis}< 0$, therefore the reversible limit is the case where the
fraction of violating trajectories is maximal.  This may look rather
unexpected as transient violations might be thought to be a
characteristic of non-equilibrium processes.

Transient violations of the second law are expected to decrease fast as
the average value of the dissipated work $\overline{W_{\rm dis}}$
increases (for instance, if the size of the system increases) becoming
unobservable in the thermodynamic limit.  How transient violations are
suppressed as the system size increases can be understood also from the
JE. We rewrite \eq{ft5} as,
\be
1=\overline{\exp\Bigl
(-\frac{W_{\rm dis}}{k_BT}\Bigr)}=P_+\Biggl[ \exp\Bigl
(-\frac{W_{\rm dis}}{k_BT}\Bigr)\Biggr]_+ + P_-\Biggl[ \exp\Bigl
(-\frac{W_{\rm dis}}{k_BT}\Bigr)\Biggr]_-
\label{ft9}
\ee
where $P_+,P_-$ are the probabilities to generate a trajectory with
$W_{\rm diss}>0,W_{\rm dis}<0$ respectively, i.e. $P_++P_-=1$. In
analogous way, the square brackets $[..]_+, [..]_-$ denote averages
$\overline{(..)}$ but restricted over the subsets of trajectories with
$W_{\rm dis}>0,W_{\rm dis}<0$ respectively. In Fig.~\ref{fig.pwfluct}
$P_+,P_-$ would correspond to the green and white areas under the
distribution. Using this decomposition it is easy to understand how work
trajectory values contribute to the r.h.s. of\eq{ft9} enforcing the
validity of the JE. The value of $W_{\rm dis}$ increases proportionally
to the size $N$ of the system (because the work is an extensive
quantity). Because all four quantities appearing in the
r.h.s. of\eq{ft9} are positive, to impose a sum of both terms of 1, the
factor $p_-$ has to be exponentially small with the average value of the
dissipated work (i.e. the size of the system) $p_-\sim \exp(-{\cal
O}(N))$ to compensate the divergence of the corresponding average
$[..]_-$. This implies $p_+=1-\exp(-{\cal O}(N))$ so transient
violations are exponentially suppressed with the system size, yet they
have to be weighed for the JE to hold.

\subsection{Free energy recovery from non-equilibrium experiments}
\label{free}
An important consequence of the JE\eq{ft4} is that non-equilibrium experiments
can be used to recover equilibrium free-energy differences~\cite{Jarzynski97a},
\be
\Delta F=-k_BT\log\Biggl(\overline{\exp\bigl(-\frac{W}{k_BT}\bigr)}\Biggr)~~~~.
\label{ft10}
\ee
The non-equilibrium work relation\eq{ft10} is useful to find the equilibrium free-energy change
along a given reaction when it is not possible to carry it out 
reversibly.  The idea is to repeat non-equilibrium experiments many
times and evaluate the exponential average in the r.h.s of\eq{ft10} to
derive the corresponding work in a reversible process. This formula
has been used to recover the free-energy change in the
folding-unfolding reaction for small RNA molecules, see
Sec.~\ref{RNAexp} for a detailed exposition.  However, there are
practical difficulties in the applicability of\eq{ft10} as the number
of trajectories included in the exponential average must be actually
infinite. This is unrealizable in practice as non-equilibrium
experiments can be performed only a finite number of times and the
finiteness of the number of trajectories introduces a bias. In what
follows we will use $\Delta F_{JE}$ to denote the estimate for the
equilibrium free energy $\Delta F$ obtained by using the JE given
in\eq{ft10} with a finite number of trajectories.  As we remarked in
the previous paragraph, the number of trajectories required to
evaluate the JE grows exponentially with the average value of the
dissipated work. The dependence of the bias and error with the number
of pulls has been estimated in some cases \cite{ZucWoo02,GorRitBus03}. In general
this dependence can be quite complicated as it depends on the behavior
of the left tails of the distribution $P(W)$ which are difficult to
analyze in general.

If the non-equilibrium experiment is done in the near-equilibrium regime
then it is better to use the fluctuation-dissipation (FD) estimate\eq{ft8},
\be
\Delta F_{FD}=\overline{W}-\frac{\sigma^2}{2k_BT}~~~.
\label{ft11}
\ee
In most cases it is difficult to determine whether the non-equilibrium
process is done in the near equilibrium regime, so this estimate has
to be taken with caution. A description of the bias and error for the
estimates\eqq{ft10}{ft11} in the near-equilibrium case has been
recently given. Both are exact in the limit of infinite number of
non-equilibrium trajectories. Interestingly, when the number of
repeated experiments is small the JE estimate\eq{ft10} provides a
better estimate than the FD\eq{ft11} does~\cite{GorRitBus03}.

\section{Experimental observation of work fluctuations}
\label{expwork}
For sake of clarity we discuss now some examples where work fluctuations
are experimentally measurable. Some of them have been already measured,
others might be in the near future. 

We start this tour discussing recent experiments on simple systems. This
is the case of a micron-sized polystyrene bead confined in an optical
trap and dragged through a solvent (e.g. water) of viscosity $\eta$ at
constant speed $v$. In average the viscous drag on the bead exerts a
force $\gamma v$ that counteracts the force inside the trap
$f(t)=-kx(t)$ where $t$ denotes the time, $k$ is the stiffness of the
trap and $x(t)$ is the distance of the bead to the center of the
trap. The control parameter is the position of the center of the trap
$x_0(t)$ that moves at a constant speed $\dot{x}_0=v$ and the
fluctuating variable is the position $x(t)$ of the bead inside the trap
(or equivalently the force $f(t)$ acting on the bead). The work along a
trajectory of duration $t_f$ is given by $W=\int_0^{t_f} dsf(s) vds$.
For such case work fluctuations were theoretically
predicted~\cite{MazJar99} and recently measured
\cite{WanSevMitSeaEva02}.

Moving to more complex systems, recent experiments have studied the
response of\break biomolecules to mechanical
force~\cite{BusSmiLipSmi00,BusMacWui00,Bao02}. The advent of
nanotechnologies has opened the possibility to exert very small forces
on nanosized systems (from piconewtons $1pN=10^{-12}N$ using optical
or magnetic tweezers, to nanonewtons $1nN=10^{9}N$ using AFMs). These
techniques allow researchers to manipulate and study individual biomolecules one
by one.  Work fluctuations
have been already observed in the unfolding of small RNA molecules
(around 100 pair bases) under the action of mechanical force. More
will be said below in Sec.~\ref{sm}. In these experiments the RNA
molecule is held through linker polymers to two micron-sized
beads. One is held by suction on the tip of a micropipette, the other
is confined in the optical trap. The molecule is pulled as the
distance $x(t)$ between the center of the optical trap and the tip of
the micropipette increases at a given rate. $x(t)$ therefore defines
the control parameter. The fluctuating variable in this case is the
force $f(t)$ exerted on the whole system that is measured through the
deflection of the bead in the trap. The work along a given trajectory
is again $W=\int_0^{t_f}f(s)\dot{x}(s)ds$~\footnote{In these experiments
usually $x(t)$ is the distance measured from the center of the bead in
the trap (rather than the center of the trap) to the center of the
bead in the micropipette. This introduces a correction to the work
that is negligible in most cases~\cite{Oregon}.}. Work fluctuations are
observed during the unfolding process due to the stochastic behavior
of the breakage force at which the molecule unfolds. Similar
experiments are expected to be conducted also for proteins~\cite{CarObeFowMarBroClaFer99}, albeit the
large molecular weight of such molecules might render the quantitative
evaluation of work fluctuations difficult.

One might speculate also on the importance of work fluctuations on the
behavior of molecular motors and their efficiency. In this case, work
fluctuations are observable through single molecule experiments by
measuring the mechanical force exerted upon the motor as it translocates
along the template. Examples of these motors are
DNA~\cite{WuiSmiYouKelBus00,GoeAstHer03} or
RNA~\cite{GelLan98,WanSchYinLanGelBlo98,ForIzhWooWuiBus02} polymerases
during the replication and transcription process, helicases and
topoisomerases that unwind the DNA~\cite{HaRasCheBabGauLohChu02} or the
ribosome during the transcription process. Other cases include the
condensation process of DNA inside the viral capside during the
infection cycle~\cite{SmiTanSmiGriAndBus01}, and gene regulatory
mechanisms (such as transcription factors) ruled by protein-DNA
interactions that expose large segments of condensed DNA to the
replication machinery~\cite{LegRobBouChaMar98}. Work fluctuations are
predicted to be observable in all these systems. Quantitative
investigations will be surely conducted in the future.

Despite of their inherent interest, the measurement of work
fluctuations in biomolecules has two important drawbacks: accuracy and
reproducibility. Indeed, few single molecule experiments are fully
reproducible due to the complexity of conditions and external factors
required. Reproducibility at the single molecule level is specially
serious in biomolecular processes requiring protein activity as many
external factors strongly affect the outcome of the experiment.
Accuracy is also an issue specially for measurements with nanometer
resolution where stability and drift of the machines (e.g. optical
tweezers) still impede high accuracy results.

Accurate and reproducible measurements of work fluctuations might be
easier in systems with reduced complexity within the traditional domain
of physics. One example is the already
mentioned experiment of the bead in an optical trap moved through a
solvent. High accuracy recent experimental measurements of the work
between non-equilibrium steady-states confirm that such measurements
are indeed possible~\cite{TreJarRitCrooBusLip03}. Another example that
has called our attention recently is the case of magnetic nanoparticle
systems in a magnetic field~\cite{Ritort03b}.  Magnetic measurements in
microsquids in Grenoble (France) have shown how it is
possible to observe magnetization reversal of single magnetic
nanoparticles of magnetic moment $\mu\sim 1000\mu_B$ ($\mu_B$ is the
Bohr magneton) at low temperatures~\cite{WerBonHasBenBarDemLoiPasMai97}. For magnetic nanoparticle systems
the control parameter is the external magnetic field that can be
switched at a constant speed (ramping experiments). The fluctuating or
stochastic variable is the value of the field at which the
magnetization reverses.  The work along a given trajectory is then
given by $W=-\mu\int_0^{t_f}M(t)\dot{H}(t)dt$.  The aspect that makes
these systems specially interesting is the possibility to use SQUID
quantum (i.e. high-precision) technology to measure the magnetic
moment in favor of a higher accuracy. Reproducibility is also easier
to achieve as many physical properties of nanoparticles can be
externally tuned, for example the height of the activation barrier and
consequently the relaxation time of the nanoparticle as well. Other specific
properties of magnetic nanoparticle systems makes them specially
suitable to measure work fluctuations \cite{Ritort03b}. We
may see these experiments done in the near future.

\subsection{Measurement of heat fluctuations}
\label{heat}
Up to now we discussed about work fluctuations but nothing was said
about heat fluctuations. The reason is simple. Work is much easier to
measure than heat. Although transferred heat can be measured by using a
small thermometer probe (by recording its
change of temperature) heat fluctuations are another matter. The easiest
procedure to measure heat along a trajectory is to use the first law of
thermodynamics where $\Delta Q=\Delta E-W$. Knowledge of both the work and the
energy change along a trajectory immediately gives the heat exchanged
between system and the bath. Measuring the
energy content of the system can be hopeless in many cases. Only in some
special cases this is possible. Here I discuss two possible situations.

The first one corresponds to the case where no energy change occurs
between the initial and final configuration for all non-equilibrium trajectories. This
situation is realized in the magnetic example~\cite{Ritort03b} discussed in the
previous section where the
reversal symmetry of the system under a field induces a zero energy
change $E_f=E_i=-\mu H_0$ if the field is changed from $-H_0$ to $H_0$
in a ramping experiment (here we assume $H_0$ to be large enough for
the initial and final magnetization to align in the direction of the
field). In general, $\Delta E=0$ can be
accomplished in any non-equilibrium cycle assuming that
the initial and final states are identical. In the case of the
unfolding of the RNA molecule under applied force this can be achieved
by considering non-equilibrium trajectories where the molecule first
unfolds and then refolds along a given cycle.

The second situation corresponds to the case where, due to the
inherent simplicity of the system, the energy is known. A relevant example
is the particle confined in an optical trap. In that case the energy of
the bead in the trap is well approximated by $E=(1/2)kx^2$ and
therefore the value of $\Delta E$ is known for each trajectory. The
distribution of exchanged heat shows interesting features as compared to
the work that have been recently discussed by Zohn and Cohen~\cite{ZohCoh03}.

\section{Single molecule experiments}
\label{sm}
The advent of nanotechnologies has provided instruments and tools for
scientists to manipulate individual molecules and follow their
dynamical trajectories as they carry out specialized molecular
tasks~\cite{BaiWanSunWol99}. The research of molecular reactions
performed by individual molecules offers new insight on the importance
of fluctuations and stochasticity in small systems.  Mechanical force
has been recognized as essential to understand the fate of many
chemical reactions~\cite{TinBus02,BusCheForIzh03}. Several
force-microscopies are currently available to investigate the
individual behavior of biomolecular complexes. Atomic force microscopy
(AFM), optical and magnetic tweezers tweezers have become common tools
that allow scientists to measure the response of these systems to applied
external force. These techniques cover different but overlapping
ranges of forces: AFM covers a range of forces spanning from several
tens of pN up to hundreds of pN, optical tweezers span the
intermediate region between 1pN and 100pN and magnetic tweezers are
sensitive to tenths of pN.

%

%
%
\begin{figure}
\begin{center}
\epsfig{file=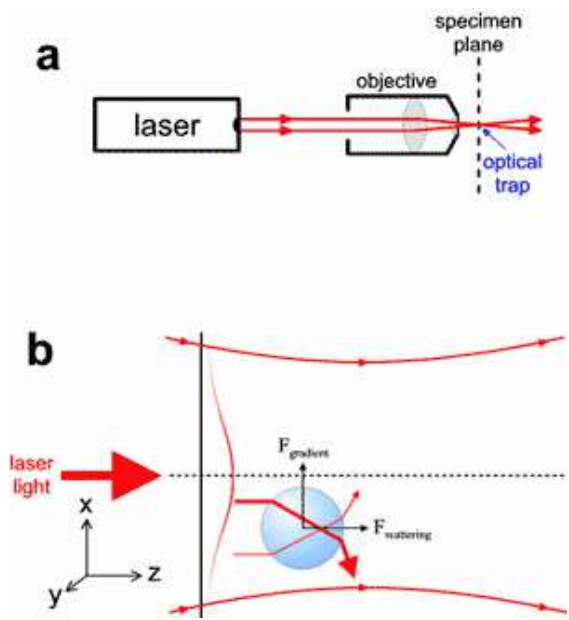,angle=0, scale=0.7}
\end{center}
\caption{A single laser tweezers setup. (a) The laser light is focused into a spot by using an objective.
(b) A Gaussian profile of light intensity generates a  confining potential due to conservation of light
momentum. The trapping force is induced by the difference in the
index of refraction between the polystyrene bead and the surrounding water.} 
\label{fig.ot}
\end{figure}

Thermal fluctuations are important whenever the energies involved in
molecular processes are of the order of several $k_BT$. A quick estimate
of the forces participating in this regime can be obtained as
follows. The typical distance $d$ involving conformational changes at
the biomolecular level is of the order of 1nm~\footnote{This is only a
rough estimate, for instance the base pair distance in DNA is around one
third of a nanometer. This is the minimal distance that polymerases have
to cover to elongate one base pair the newly synthesized strand.}. At
room temperature $T=298K$, $Fd=k_BT$ this gives a force $F\simeq
4pN$. Optical tweezers are ideal to investigate a large region of
intermediate forces around this value~\cite{SmiCuiBus02}.  Nowadays, optical tweezers are
used to investigate many processes operated by biomolecules, ranging
from the elastic deformation of nucleic acids or proteins to the
specific action of enzymes acting on molecular substrates~\cite{LanBlo03}. A typical
experimental setup is shown in Fig.~\ref{fig.ot}. Optical tweezers use light
momentum conservation to generate a force gradient on polystyrene beads
(of a diameter between 1 and 3 microns) that are immersed in
water. Light deflection inside the beads arises from the difference in
the index of refraction between the beads and water. In this way a
confining potential can be generated by focusing a beam of light inside
the chamber. To a high degree the confining potential can be considered
as harmonic.  Single beam tweezers can generate confining forces of the
order of several tens of pN~\footnote{The confining force depends on
wavelength of the light. Typical wavelength values are in the range
700-1000nm, lower frequencies are inadvisable as they can lead to light
absorption and subsequent heat convection effects around the
bead.}. Dual tweezers use two counter propagating beams to generate
higher forces (up to 150 pN) and have the advantage (by measuring the
total amount of deflected light) that recurrent calibration is not
required to measure forces.  A fluid chamber is fixed in a movable
stage or frame that is controlled by a piezo actuator. The chamber is
made out of two parallel glass plates separated by a thin layer of
parafilm. Inside the chamber there is a glass micropipette that can trap
beads of the size of the micron by air suction. The two counter
propagating laser beams can confine another bead in the optical trap. To
measure forces on molecules a tether is attached to the two beads (one
in the micropipette, the other in the trap). Attachments are designed by
chemical treatment of the surface of the beads and chemical modification
of the ends of the molecule (called labeling).  As the stage is
moved the force on the bead in the trap (and therefore, on the tether)
can be measured. The distance between beads is then measured by using a
light lever and a force-extension curve (FEC) can be recorded.
Optical tweezers have been used
in different fields ranging from physics to biology. A 
survey of their applications can be found in~\cite{LanBlo03}.

\begin{figure}
\begin{center}
\epsfig{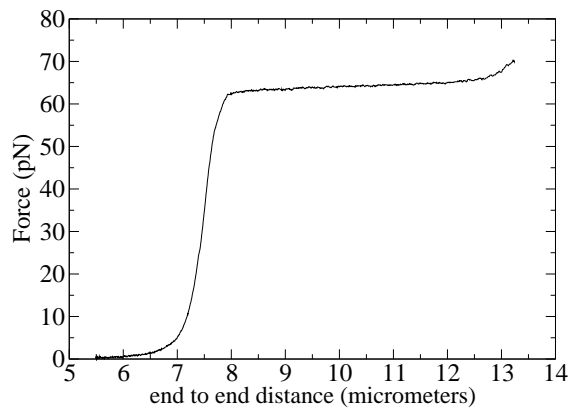}
\end{center}
\caption{Force extension curve for a torsionally unconstrained DNA molecule of the $\lambda$
bacteriophage, 24000 base pairs long in a water buffer at 100mM NaCl
concentration and 7pH. The molecule has a contour length of
approximately $8\mu m$ and shows the characteristic overstretching
transition around $65$ pN.}
\label{fig.OS}
\end{figure}

DNA plays a central role in
biophysics~\cite{FrankKamenetskii97,CalDre97}. Accordingly its
mechanical properties have been extensively investigated during the past
10 years~\cite{StrAleCroBen00,BusBrySmi03}. Initial investigations on
the elastic response of double-stranded DNA under
tension~\cite{SmiFinBus92} have revealed that DNA behaves like an
entropic spring as predicted by the worm-like chain model of polymer
theory~\cite{BusMarSigSmi94}. However, at difference with other polymers
DNA shows structural transitions at modest forces (around or below
100pN) depending on how the molecule is pulled. For example, torsionally
unconstrained double-stranded DNA shows a highly
cooperative overstretching  transition around
65pN~\cite{CluLebHelLavVioChaCar96,SmiCuiBus96}.  At the origin of this
behavior there is the double-helix structure of DNA and the
associated uncoiling of the two strands. However, if both strands are
pulled from the same end of the DNA molecule, then DNA sequentially
unzips at constant force following a curve that depends on
the particular nucleotide
sequence~\cite{EssBocHes97,DanColBouLubNelPre03}. The elastic response of
DNA has produced many
experimental~\cite{WanLanGelBlo97,LegRomSarRobBouChaMar99,RieClaGau99,WilRouBlo02,BryStoGorSmiCozBus03}
as well as theoretical
investigations~\cite{LebLav96,BouMez98,Marko98,AhsRudBru98,CocMon99,ZhoZhaYan00}
to characterize its structural transitions. A
particular force-extension curve (FEC) showing the characteristic
overstretching transition of double-stranded DNA is shown in
Fig.~\ref{fig.OS}.

\section{Pulling experiments on RNA}
\label{RNAexp}
RNA is an essential molecule in biochemistry. It plays an intermediate
role between DNA (which encodes the genetic information and represents
the ``software'' in living organisms) and proteins (which perform
specialized tasks inside the cells and represent the ``hardware''). Such
intermediate role has been emphasized after the discovery that certain
RNA molecules (called ribozymes) have catalytic activities that are
essential in many regulational processes~\cite{GesCecAtk99}. The
relevance of RNA has motivated many single molecule
studies. Compared to DNA, optical tweezers
measurements in RNA present additional difficulties to the
experimentalist.  Not only RNA requires more elaboration in the
synthesis of the molecular constructs, it is also a molecule very
sensitive to the surrounding environment and degrades easier. Moreover,
RNA domains have extensions of few tens of nanometers after unfolding,
thus requiring more careful and precise measurements.

\begin{figure}
\begin{center}
\epsfig{file=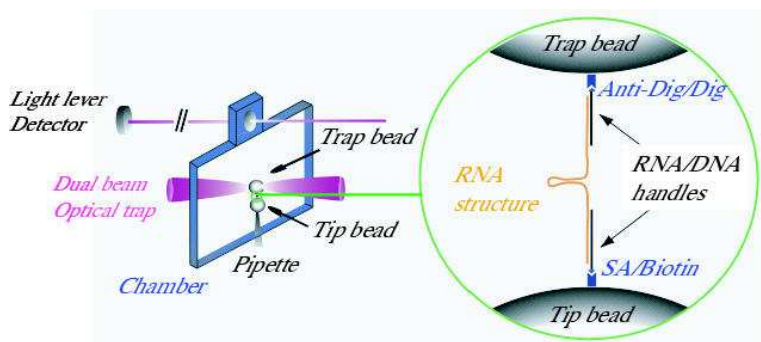,angle=0, scale=0.5}
\end{center}
\caption{Typical experimental setup when pulling RNA molecules. The
molecular construct consists of an RNA molecule attached by its ends to
two RNA/DNA hybrid handles (to avoid formation of secondary structures
in the handles). As compared to DNA, RNA single molecule experiments
present additional difficulties as RNA quickly degrades and the
resolution required to observe the unravelling of the molecule is much
higher and of the order of the nanometer.}
\label{tweezersRNA}
\end{figure}

Liphardt et al.~\cite{LipOnoSmiTinBus01} have pulled RNA molecules and
studied their unfolding by applying external force using optical
tweezers.  The molecular construct consists of two hybrid DNA-RNA
handles that are annealed to the ends of a small RNA molecule,
Fig~\ref{tweezersRNA}. As the molecular construct is pulled the
force-extension curve (FEC) reflects the elastic behavior of the
handles (well described by a worm-like chain
model~\cite{BusMarSigSmi94}) until a force is reached where the
molecule unfolds and a jump in the force and distance is observed.

\begin{figure}
\begin{center}
\epsfig{file=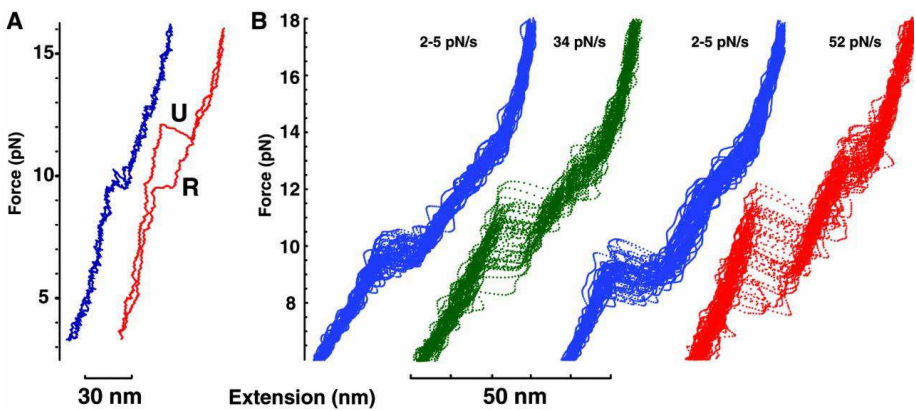,angle=0, scale=0.6}
\end{center}

\caption{Non-equilibrium pulls in the RNA molecule P5abc at different
pulling speeds in a buffer in the absence of magnesium. Panel A shows a
reversible (left blue, pulling speed equal to 3-4pN/s) trajectory and an
irreversible trajectory (right red, pulling speed 52pN/s). Panel B shows
unfolding trajectories for two pulling speeds, 34pN/s (green) and
52pN/s(red) compared to near-equilibrium pulls at 2-5pN/s (blue). Figure
taken from \cite{LipOnoSmiTinBus01}.}
\label{fig.pulls}
\end{figure}

Very interesting dynamical effects were later observed in small RNA
hairpins depending on the pulling rate. For slow pulling rates the
molecule was seen to follow always the same trajectory and unfold at a
reproducible value of the critical force~\footnote{Reproducibility of
trajectories has always limitations imposed by the unavoidable drift
of the optical tweezers machine, see the remark at the end of
Sec.~\ref{expwork}. The accuracy in the value of the breakage force
can be well controlled.}. At this force coexistence and hopping
between the folded and unfolded conformations has been observed
characteristic of cooperative
unfolding~\cite{LipOnoSmiTinBus01}~\footnote{The dependence of the
value of the transition force and the hopping frequency on the sequence of the RNA molecule has
been studied in~\cite{CocMarMon03}}. More interesting, as the
pulling rate increases larger hysteresis and stochastic fluctuations
in the value of the breakage force were observed.  Typically the
average value of the breakage force tends to increase with the pulling
rate. This dependence has been investigated by Evans and
Ritchie~\cite{EvaRit97,EvaRit99} who have applied Kramers theory
\cite{HanTalBor90} to describe the activated dynamics of a particle
jumping over a force-dependent barrier as described by
Bell~\cite{Bell78}.  The study of the loading rate dependence of the
breakage force in this type of systems has led to new developments in
what is now commonly referred as single-molecule force spectroscopy
\cite{Evans01,FriWehKuhGau03}, a technique that is useful to investigate the energy
landscape of molecular interactions.  Typical unfolding curves showing
the pulling rate dependence of the breakage force and the resulting
hysteresis effects are shown in Fig.~\ref{fig.pulls}.

\begin{figure}
\begin{center}
\epsfig{file=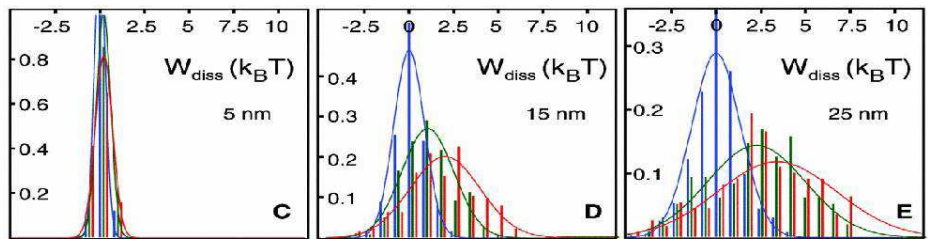,angle=0, scale=0.62}
\end{center}
\caption{Work distributions for the P5abc molecule at different pulling
speeds (3-5pN/s blue, 34pN/s green, 52pN/s red) measured at different 
distances along the pulling process. Figure
taken from \cite{LipDumSmiTinBus02}.}
\label{fig.pwjan}
\end{figure}

Hummer and Szabo have realized~\cite{HumSza01} that the
non-equilibrium work relation\eq{ft10} can be used in single molecule
experiments to reconstruct the free energy landscape along the force
coordinate. The JE\eq{ft10} has been experimentally tested
in~\cite{LipDumSmiTinBus02} for the P5abc hairpin by repeated
measurements of the work done along the unfolding trajectory at
different pulling speeds. For P5abc in EDTA buffer the unfolding free
energy change is well known from its secondary structure and therefore
is a useful example to test the validity of the JE. Typical work
histograms are shown in Fig.~\ref{fig.pwjan} for three pulling
speeds. As expected, as the pulling rate increases the average value
of the dissipated work increases reaching values of the order of
$4k_BT$ at the fastest pulling speeds. The main result in
\cite{LipDumSmiTinBus02} is that the JE can be used to predict the
free energy change for the folding-unfolding transition in the P5abc
hairpin with a precision within $1k_BT$ using a modest number of pulls
(around 100). Moreover, the JE provides a better estimate for the
equilibrium free-energy change than the FD estimate does\eq{ft11}. The
advantage of the former as compared to the later has been
verified in the near-equilibrium regime
(where\eq{ft8} holds), when the number of repeated pulls is not too
large~\cite{GorRitBus03}.

\begin{figure}
\begin{center}
\epsfig{file=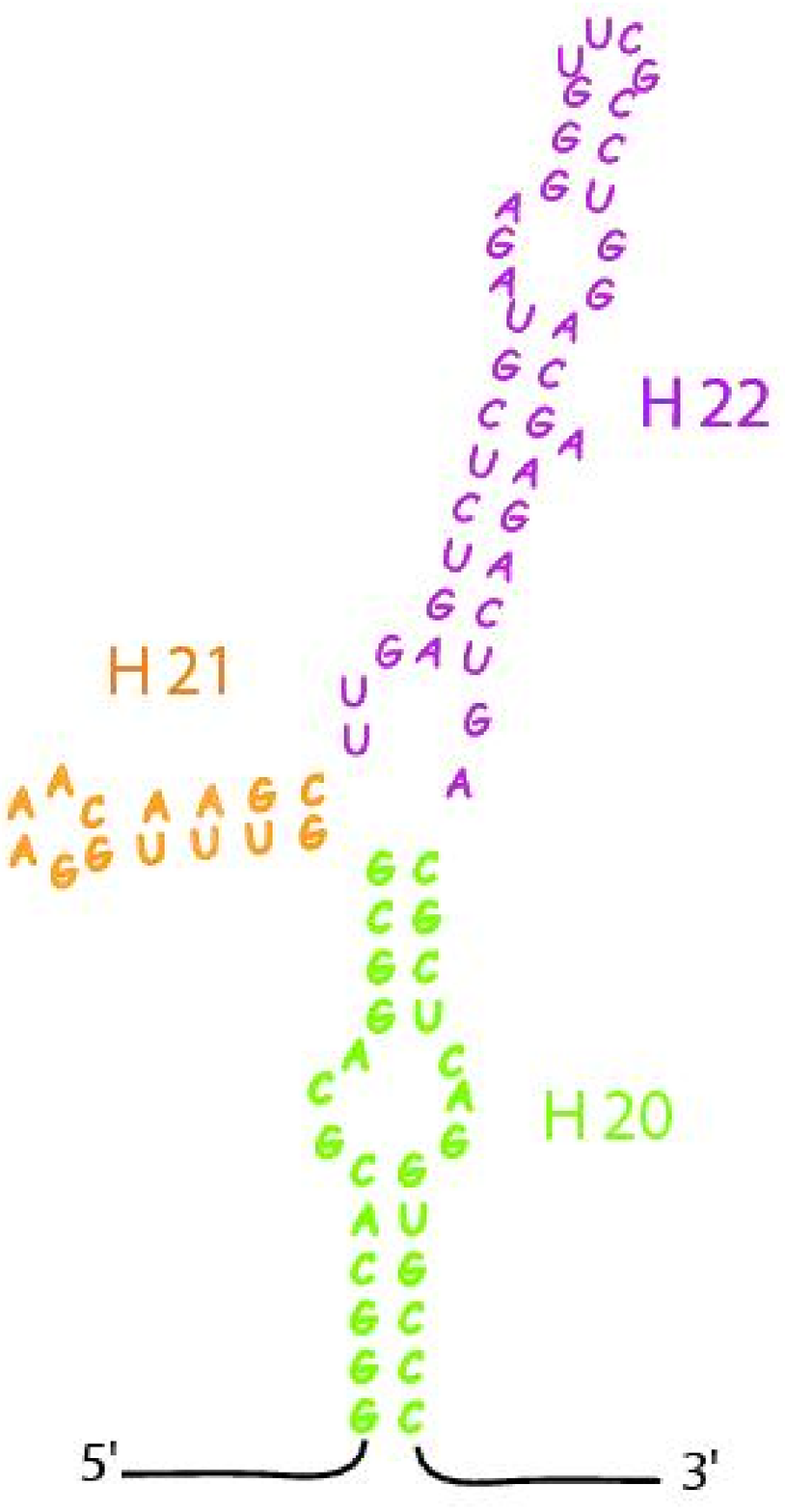,angle=0, scale=0.4}\epsfig{file=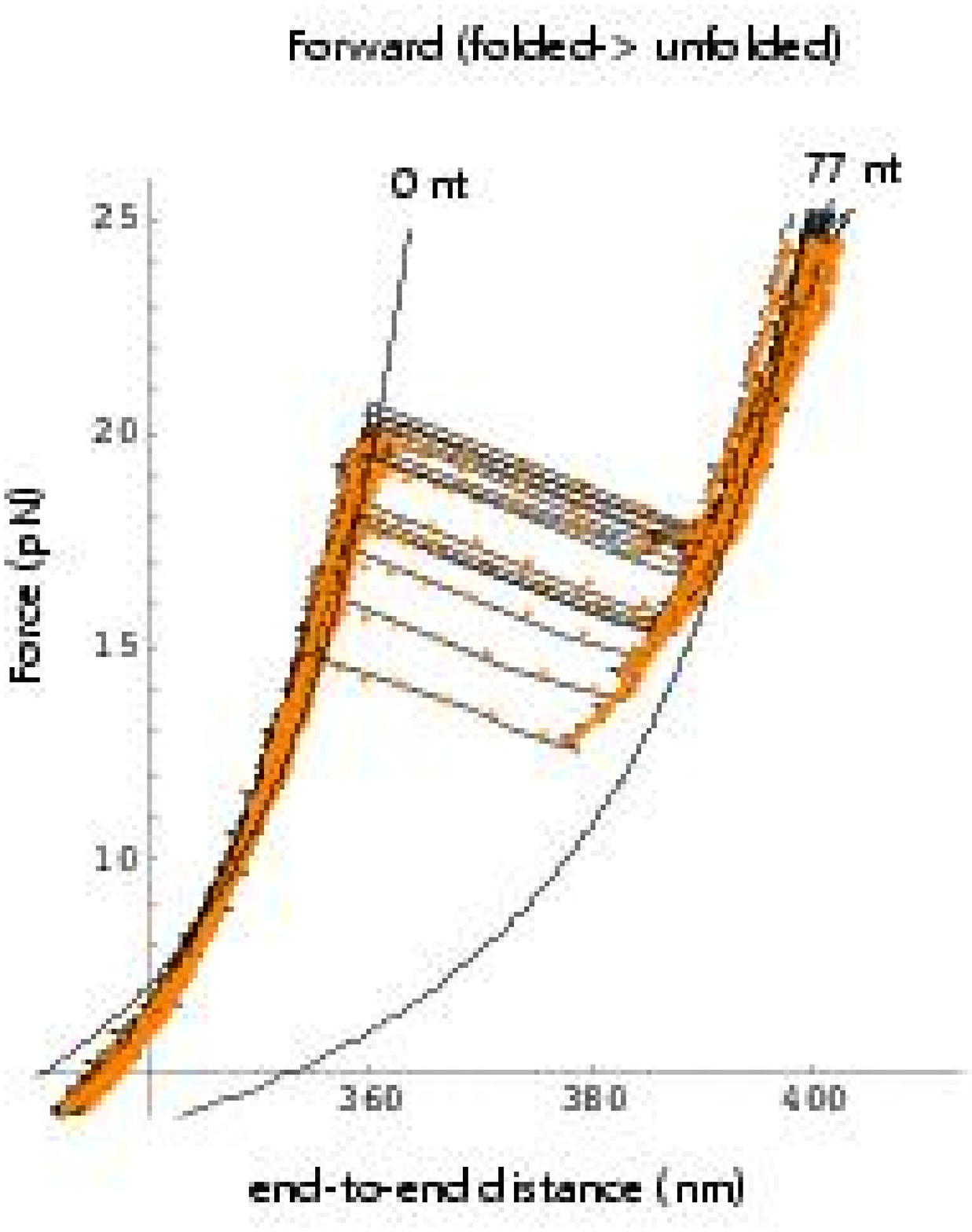,angle=0, scale=0.35}
\end{center}
\caption{The left figure shows the structure of the junction in the 16S domain of the 30S
ribosomal RNA subunit. The right figure shows some unfolding curves at
3.5pN/s. Courtesy of Delphine Collin.}
\label{fig.structure1}
\end{figure}

For the case of the P5abc in EDTA buffer the value of the dissipated
work is small~\footnote{In this buffer conditions kinetic barriers are
low. High kinetic barriers and strong irreversibility are obtained
either by going to faster pulling speeds (however, this is not easy to
accomplish due to limited experimental capabilities) or in different
buffer conditions. For the latter, high kinetic barriers and many
intermediate states are obtained in the presence of divalent cations
such as magnesium that establish specific tertiary contacts between some
bases.}. In general, for larger molecules the JE is expected to give
less reliable estimates for the equilibrium free-energy as the value of
the average dissipated work increases. An example is shown in
Fig.~\ref{fig.structure1}.  Typical unfolding curves for a three way RNA
junction are shown in Fig.~\ref{fig.structure1}. The validity of the JE
for such cases is currently investigated~\cite{RitColJarSmiTinBus03}.
Other more complex cases imply the unfolding of even larger RNA
molecules consisting of many domains such as the recently investigated
L21 RNA ribozyme~\cite{OnoDumLipSmiTinBus03}.

\section{Modeling the experiment}
\label{modexp}
\begin{figure}
\begin{center}
\epsfig{file=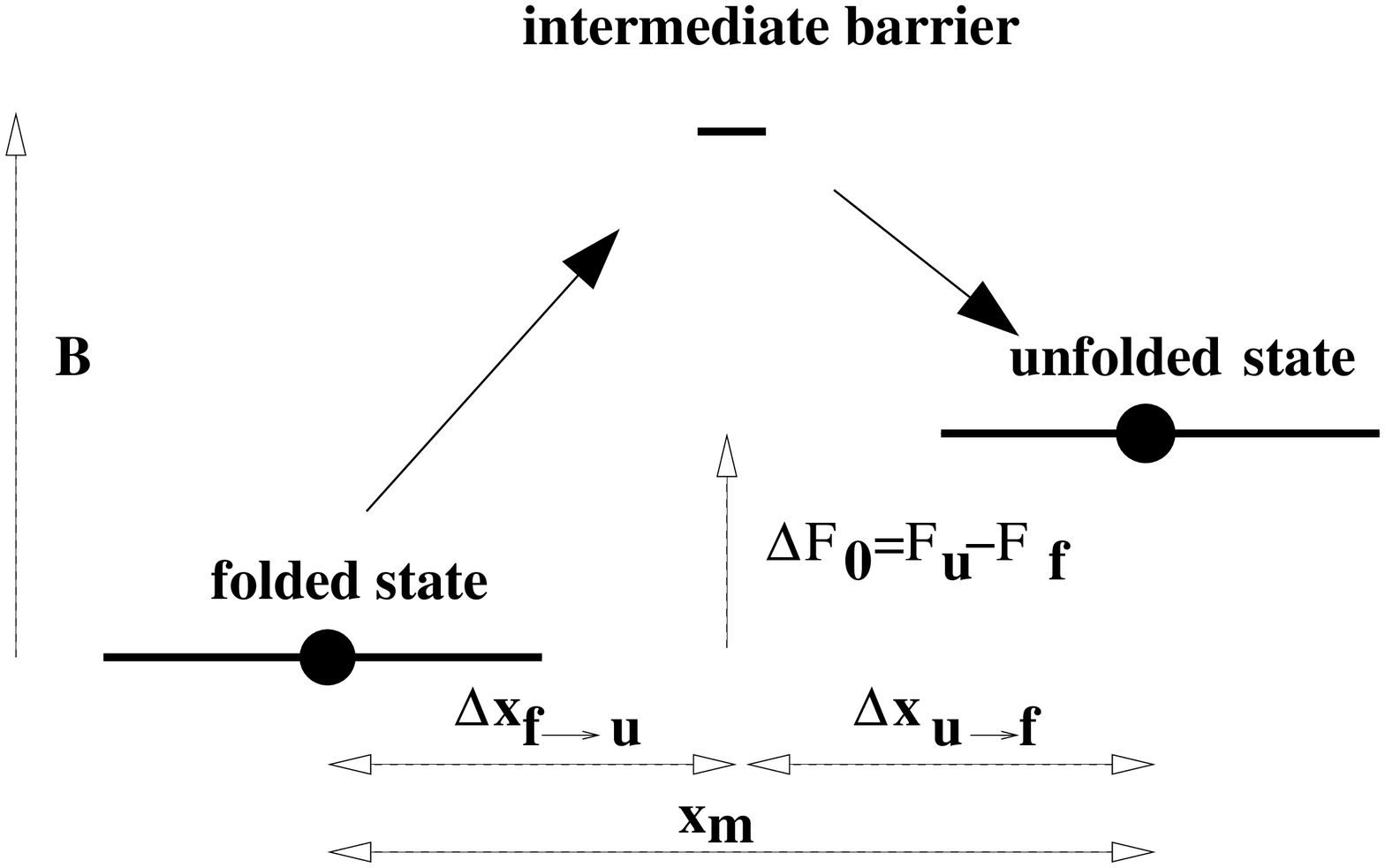,angle=0, scale=0.4}
\end{center}
\caption{The two-state model with an intermediate barrier. The
parameters are the free-energy gap $\Delta F_0$, the unfolding distance
$x_m$, the height $B$ of the intermediate barrier and the distance of
the intermediate barrier to the folded state $\Delta x_{f\to
u}=x_m-\Delta x_{u\to f}$.  Figure taken from \cite{RitBusTin02}.}
\label{fig1.PNAS}
\end{figure}

In~\cite{RitBusTin02} a two-state model has been studied to justify
the non-equilibrium experiments in \cite{LipDumSmiTinBus02}. The main
goal was to confirm that indeed it is possible to obtain the
equilibrium free energy (within an error equal to $1k_BT$) by using
the JE with a limited number of pulls done in that
experiment. Interestingly, with this model it is possible to go quite
far and find out several results regarding the kinetics of the
unfolding process. This allows to make also specific predictions about
the kinetic dependence of the dissipated work that can be
experimentally tested as well as quantitative statements about the
validity of the JE for two-state systems.  Moreover, it is possible to
do explicit calculations for the work distribution $P(W)$ and, if
desired, go beyond the Gaussian case\eq{ft7}. Two-state models provide
phenomenological descriptions of systems that can exist in two
different forms, therefore the following considerations are expected
to be applicable to many systems beyond the folding-unfolding dynamics
of RNA molecules.  In fact, the two-state model has been shown to
provide a good description of the folding-unfolding dynamics of small
DNA or RNA hairpins that display strong cooperativity
\cite{BonKriLib98,CheDil00} as well as structural transitions in
polymers~\cite{RieFerGau98}. The model
is represented in Fig.~\ref{fig1.PNAS} where the two conformations
(folded and unfolded) are separated by an intermediate barrier located
at a distance $\xfu$ from the folded state and $\xuf$ from the
unfolded state, the value $x_m=\xfu+\xuf$ being the total distance
between the folded and the unfolded states. The free energy difference
between the two states is denoted as $\Delta F_0$ and the height of
the barrier is indicated as $B$. Transition rates between the folded
and the unfolded state are thermally activated and force
dependent~\cite{Bell78},
\bea
\kfu(f)=k_m k_0 \exp(-\beta(B-f \xfu))\nonumber\\
\kuf(f)=k_m k_0\exp(-\beta(B-\dF+f \xuf))
\label{mod1}
\eea
where $f$ is the external force, $\beta=1/k_BT$, $k_0$ is a
microscopic attempt frequency and $k_m$ is a contribution arising from
the handles, the bead in the trap and the machine~\footnote{In
\eq{mod1} we continue to use the term $F$ for the Gibbs free
energy. To be precise we should use instead $G$ as for the
experimental conditions the temperature and pressure of the bath are
held constant.}. The rates\eq{mod1} satisfy detailed balance, a
necessary condition for the equilibrium regime to be characterized 
by Boltzmann populations of the folded and unfolded states.  The
dynamics of the two-state model under the action of an
external force has been analyzed in detail for the case of
no-refolding process along the unfolding
curve~\cite{EvaRit97,EvaRit99}, also called a first-order Markov
process. This particular case is analytically tractable and specific
predictions about the form of the work distribution can be made~\cite{ManRit03}.

In the theoretical treatment of a non-equilibrium pulling experiment the
force can be taken as the control parameter~\footnote{Strictly speaking
this is not true. As remarked in Sec.~\ref{expwork} the control
parameter in pulling experiments using optical tweezers is not the force
but the distance between the center of the optical trap and the tip of
the micropipette. The pulling speed $r$ is always an average value of
the force-dependent speed along the unfolding curve. Under this
approximation (which typically introduces a small correction), using the
force or the distance as the control parameter turns out to be
equivalent as $d(fx)=fdx+xdf$, see footnote (**) in \cite{RitBusTin02} for a
related remark.} and
increased at an approximately constant rate $r=\dot{f}$. The work
exerted along a given trajectory is taken as $W=\int xdf$. The
probability distribution cannot be exactly evaluated in closed form 
and only the moments of the distribution can be computed in
a perturbative scheme where the average dissipated work is assumed to be several times $k_BT$. The
results have been given in~\cite{RitBusTin02} for the first two moments. These give the
average dissipated work and its variance, from which the
value of the fluctuation-dissipation ratio $R$ can be inferred. The first two moments are the most
relevant quantities as they can be directly compared with the
experimental results.
A general result
for the average dissipated work can be derived in the linear-response
regime where the pulling speed is slow compared to the hopping frequency
at the transition force $f_t$ (i.e. the value of the force at which
the folded and unfolded populations of the RNA molecules are equal in
equilibrium). The linear-response regime is therefore characterized by
the dimensionless parameter $\rho$ defined as,
\be
\rho=\frac{r}{f_tk_{\rm total}(f_t)}
\label{mod2}
\ee
where $k_{\rm total}(f_t)=\kuf(f_t)+\kfu(f_t)$ is the total rate at the
transition force. When $\rho< 1$ the average dissipated work is given
by,
\be
\overline{W_{\rm dis}}\sim \rho\Delta F_0+{\cal O}(\rho^2) 
\label{mod3}
\ee
\begin{figure}
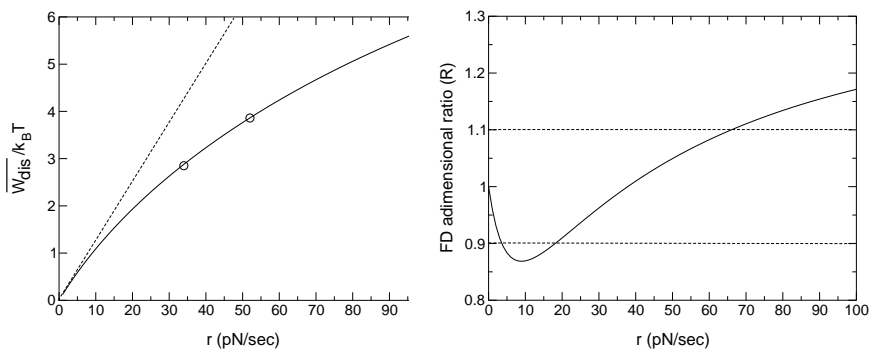

\begin{center}
\vspace{.2cm}\epsfig{file=fig2.PNAS.eps,angle=0, scale=0.25}\epsfig{file=fig3.PNAS.eps,angle=0, scale=0.25}
\end{center}
\caption{Average dissipated work (left) and fluctuation-dissipation
ratio $R$ (right) as a function of the pulling speed. The circles in
the left figure are the experimental values. The values of the kinetic
parameters characterizing the rates\eq{mod1} have been chosen to fit
the experimental data.  The dashed line in the left figure shows the
linear-response formula\eq{mod3}. The two horizontal dashed lines in
the right figure limit a region of pulling speeds where the FD
estimate\eq{ft11} is expected to approximate well the
free-energy change during the folding-unfolding reaction.  Figure taken from \cite{RitBusTin02}.}
\label{fig23.PNAS}
\end{figure}

\begin{figure}
\begin{center}
\vspace{.3cm}
\epsfig{file=fig4.PNAS.eps,angle=0, scale=0.27}\epsfig{file=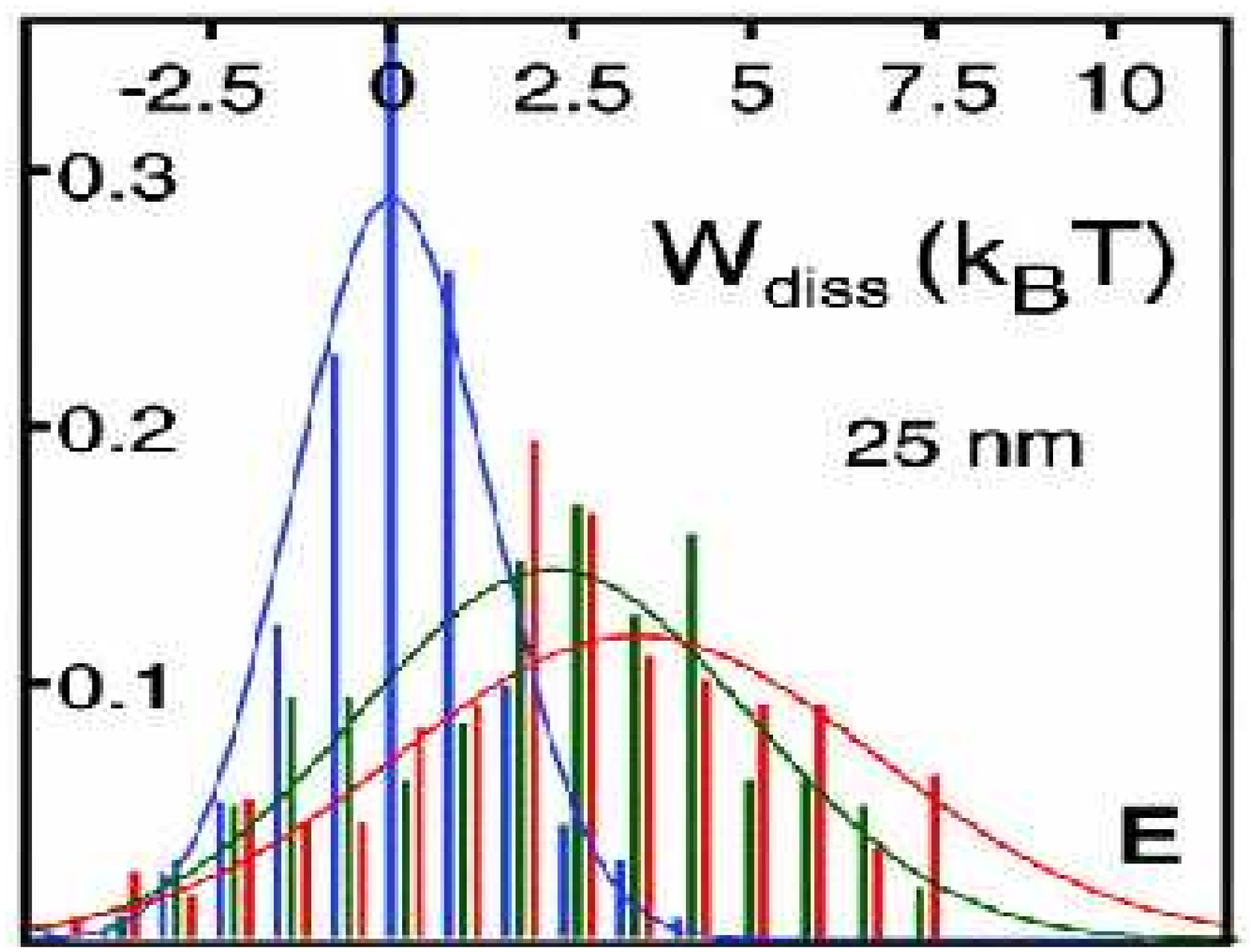,angle=0, scale=0.3}
\end{center}
\caption{Work distributions obtained for the model (left) compared to
the experimental results. Strong deviations from a Gaussian behavior
are predicted, specially in the left tails of the
distribution. Figures taken from
\cite{RitBusTin02,LipDumSmiTinBus02}.}
\label{fig4.PNAS}
\end{figure}

The linear dependence of\eq{mod3} can be used to derive estimates for
the relaxation time of the molecule that might complement other type
of kinetic measurements (such as the measurement the folding-unfolding
hopping frequency right at the transition force).  In
Fig.~\ref{fig23.PNAS} we show the results for these quantities as a
function of the pulling rate using some kinetic parameters in\eq{mod1}
to fit the experimental data. The dashed line in the left
panel of Fig.~\ref{fig23.PNAS} shows the linear response
prediction\eq{mod3}. Note that the experimental points fall off the
linear response curve, showing that the pulling rates investigated
in~\cite{OnoDumLipSmiTinBus03} explore the far from equilibrium
regime. This is an important result because it shows that, despite of
the smallness of the value of the average dissipated work (in the
range $2-4k_BT$) the experiments were carried out far from equilibrium
reinforcing the validity of the Jarzynski relation in such
regime~\footnote{Indeed, had the experiments been carried out in the
near-equilibrium regime, then the recovery of the equilibrium
free-energy change using the JE would be expected due to the
smallness of the average values of the dissipated work. The fact that
the value of the dissipated work is small, yet the system is far from
equilibrium, is consequence of the smallness of the RNA molecule.}.  This conclusion is substantiated by the dependence of
the fluctuation-dissipation ratio (right panel in
Fig.~\ref{fig23.PNAS}) which shows a strong non-monotonic behavior for
pulling speeds above 20pN/s.  Further evidence endorsing the fact that
experiments were carried out far from the equilibrium regime is
inferred from the shape of the work probability distributions
$P(W)$~\footnote{It must be emphasized though that this statement
would not be valid if, by some reason, the parameters used to fit the
kinetic data were completely off from the actual values. Although
far-fetched, this possibility cannot be ruled out as the two-state
model here considered is probably a crude approximation to the real
description of the unfolding process (see the discussion in
Ref.~\cite{BusCheForIzh03}). New experiments in other RNA hairpins are
required to reach a better understanding.}. The
results are shown in the left panel of Fig.~\ref{fig4.PNAS} and were
obtained from numerical simulations of the model using the kinetic
values of the fitting parameters as derived from
Fig.~\ref{fig23.PNAS}. Gaussian behavior is a fingerprint of the
near-equilibrium regime, see the discussion in the paragraph
containing footnote~\ref{Gauss} in Sec.~\ref{ft}. As a comparison we show in the right panel of
Fig.~\ref{fig4.PNAS} the histograms obtained from the
experiments. Both theory and experiments bear a close resemblance.
Fig.~\ref{fig4.PNAS} reveals the existence of long tails at both sides
of the work distribution that strongly deviate from the Gaussian
behavior.

\begin{figure}
\begin{center}
\epsfig{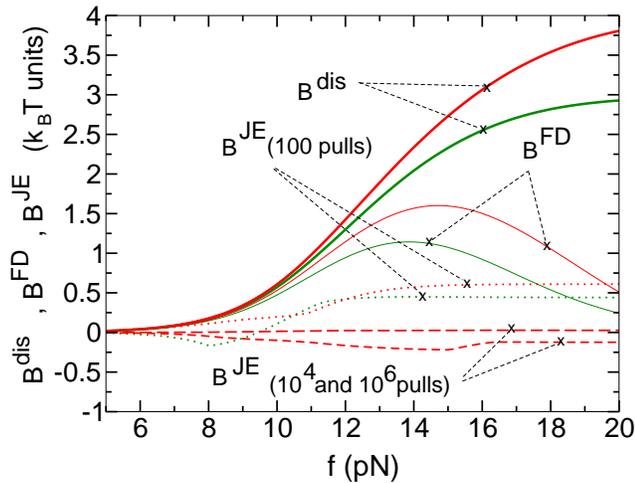}
\end{center}
\caption{Behavior of the different bias defined in
\eqqq{mod4a}{mod4b}{mod4c} along the force coordinate.
Figure taken from \cite{RitBusTin02}.}
\label{fig5.PNAS}
\end{figure}

Finally we provide an answer to the original question with which we
started this section. Can we support the main result of the
experiment~\cite{LipDumSmiTinBus02} where a small number of pulls (around 60) was enough to obtain the
equilibrium free energy (within an error of $1k_BT$) by using the JE? 
In Fig.~\ref{fig5.PNAS} we compare the bias error obtained from the two estimates\eqq{ft10}{ft11}
as well as from the average dissipated work $\overline{W}_{\rm dis}$ along the force
coordinate. The different bias errors are defined as,
\bea
B^{\rm dis}=\overline{W}-\Delta F=\overline{W}_{\rm dis}\label{mod4a}\\
B^{FD}=\Delta F_{FD}-\Delta F=\overline{W}_{\rm dis}(1-R)\label{mod4b}\\
B^{JE}=\Delta F_{JE}-\Delta
F=-\log\Bigl(\overline{\exp(-\frac{W_{\rm
dis}}{k_BT})}\Bigr)\label{mod4c}
\eea
where \eq{ft8} and \eqq{ft10}{ft11} have been used.  The bias error
depends on the number of pulls $N_{\rm pulls}$. To
obtain these bias values we have averaged \eqqq{mod4a}{mod4b}{mod4c}
over a large number of sets of experiments, each set characterized by
$N_{\rm pulls}$ repeated pulls. Full convergence to the correct
free-energy, as the number of pulls increases, corresponds to a
vanishing bias throughout the force axis. From Fig.~\ref{fig5.PNAS}
we can see how the values obtained from the average dissipated
work\eq{mod4a} and the FD estimate\eq{mod4b} quickly converge to
limiting curves characterized by a finite bias. However, the bias
obtained from the JE\eq{mod4c} slowly converges to zero and
practically vanishes only for $N_{\rm pulls}\sim 10^6$. Also, from the
JE bias\eq{mod4c} shown in Fig.~\ref{fig5.PNAS} we learn that 100
pulls are enough to get an estimate of the free-energy within $1k_BT$
of error for the folding-unfolding reaction, by using non-equilibrium work
values at the two largest pulling speeds. We mention also that the FD
estimate works well in the large force region of the force axis
(around 20pN) but not in the intermediate force region (around 14pN)
where it develops a bump. The reason why the FD works so well at high
forces has it root in the behavior of the fluctuation-dissipation
ratio $R$ shown in the right panel in Fig.~\ref{fig23.PNAS}. There $R$
has been evaluated from a pulling protocol where the force is ramped
from 0 to a value around 20pN. In that case $R$ is close to 1 for a
large region of pulling speeds (delimited by the two horizontal dashed
lines). Whenever $R\sim 1$ the FD estimate is expected to work well if
the number of pulls is not too small.

\begin{figure}
\begin{center}
\epsfig{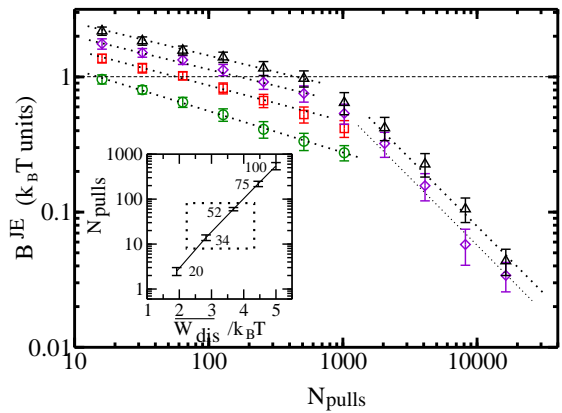}
\end{center}
\caption{Main panel: Bias error (in units of $k_BT$) for the Jarzynski
average \eq{mod4c} as function of the number of pulls for different pulling rates
(from bottom to top: 34 (green),52 (red),75 (violet),100 (black)  pN/s). Data have been
averaged over 1000 sets and error bars correspond to 100 sets. 
 Inset: Number of pulls necessary to obtain an estimate for the equilibrium free energy
within $k_BT$ and fit to the estimate $N_{\rm pulls}\sim \exp(R_-\frac{\overline{W_{\rm dis}}}{k_BT})$
which yields $R_-\simeq 1.5$.}
\label{fig6.PNAS}
\end{figure}

The number of pulls required to obtain the equilibrium free energy
with an error within $1k_BT$ can be estimated by measuring $B^{JE}$ averaged over
many sets, each one containing $N_{\rm pulls}$ repeated pulls. The
dependence of the bias $B^{JE}$ with $N_{\rm pulls}$ is shown in
Fig.~\ref{fig6.PNAS}. The decay of the bias with $N_{\rm pulls}$ can
be very well approximated by a power law $(N_{\rm
pulls})^{-\alpha(r)}$ where the exponent $\alpha(r)$ depends on the
pulling rate (or the average dissipated work as they are related each
other). The bias $B^{JE}$ shows as a crossover to a $1/N_{\rm pulls}$
behavior for $N_{\rm pulls}> 1000$ in agreement with the prediction by
Wood \cite{Wood91}. For a Gaussian process in the near-equilibrium
regime the value of the exponent $\alpha(r)$ has been estimated
numerically \cite{GorRitBus03} and is relatively close to the values
found in this case.  From Fig.~\ref{fig6.PNAS} we see the number
of pulls required for the bias $B^{JE}$ to be equal to $1k_BT$
(indicated as the horizontal dashed line). This number of pulls is
then shown in the inset of Fig.~\ref{fig6.PNAS} as a function of the
average dissipated work (also the pulling speeds are indicated). Under
certain assumptions (see \cite{Ritort04}), and only for modest values of the average
dissipated work~\cite{Ritort04}, this number of pulls can be shown to
approximately grow as $\exp(R_-\frac{\overline{W_{\rm dis}}}{k_BT})$
with $R_-$ a constant of order unity that characterizes the left side
tail of the work distribution.  For the experimental values of the pulling speed considered in
the experiment (the region limited by the square box shown in the
figure in the Inset) the required number of pulls to get the desired
accuracy is of the order of several tens as done in the experiment.

All in all, the two-state model reproduces quantitatively many aspects of the
non-equilibrium behavior observed in the experiment \cite{OnoDumLipSmiTinBus03} and justifies the
test of the validity of the JE there claimed.

\section{Conclusions}
\label{conc}
Thermodynamics represented a great step in the development of
science. It provided a general framework to understand all natural
processes that involve the transformation of different sorts of energy
(mechanical, chemical, electromagnetic) into work and heat.  While work
can be viewed as useful energy, heat represents energy that is not
useful. The second law of thermodynamics limits the amount of useful
work that can be extracted from heat. As heat abounds in nature it seems
plausible that the level of organization that we see today in the form
of biological matter originates from certain properties that
characterize heat exchange processes.  

Statistical mechanics provided a
mechanistic picture of the abstract concepts of thermodynamics in terms
of the average behavior of a large number of atoms or molecules and
their interactions. According to this picture, thermodynamic quantities
are not strictly constant but fluctuate around their average
values. However, the amount of these fluctuations is small relative to
the value of the thermodynamic quantities themselves. Much larger
fluctuations are hardly observable and become irrelevant as the
macroscopic level is approached.

Fluctuation theorems go beyond this statistical level of description by
quantifying fluctuations arbitrarily large whose magnitude can be of the
same order of the average value. This is the content of the
non-equilibrium work relation originally derived by
Jarzynski~\footnote{We did not mention in this feature extensions of the
classical non-equilibrium work relation to the quantum regime. Although
several papers have recently appeared in the
literature~\cite{Kurchan00,Tasaki00,Yukawa00,Mukamel03}, the concept of
a quantum trajectory and quantum work are to be clarified and the first
experimental attempt to test the corresponding quantum relations is
still to be done.}. In that case, work trajectories quite far from the
average or most probable trajectory, have to be properly weighed for the
equality to be satisfied. As the system size increases (or as the time
increases for steady state systems) the probability to observe these
rare trajectories quickly decreases. Were we repeat many times the
dynamical experiment, the time we should wait until finding a trajectory
that notably reduces the bias error associated with the equality,
increases exponentially with the size of the system, ultimately reaching
values that are of the order of the Poincare recurrence time. We then
considered the suggestive fact that most of the non-equilibrium
trajectories that enforce the validity of the Jarzynski equality, are
also those that inspired many of the paradoxes underlying the
statistical interpretation of heat and that were proposed in the early
days of statistical mechanics.

What is the fundamental value of these rare trajectories described by
fluctuation-theorems? If fluctuation theorems were just theorems, the
value would be predominantly academic. There is of course interest in
using the non-equilibrium work relation to obtain free energies for
transformations that cannot be carried out reversibly.  However, it
might be possible that fluctuation theorems have an added fundamental
value. They could provide physicists with a tool to explore the validity
of the principles underlying some energy transformation processes. In
the same way that classical mechanics proved inadequate to describe
energy exchange between radiation and matter at the atomistic level, one
could imagine that current theories describing thermal exchange
processes occurring at very small length-scales or short times should be
accordingly revised. In fact, all fluctuation theorems use in one way or
another the concept of microscopic reversibility. This condition ensures
that systems thermalize if left to evolve for a long time. However, it
might be possible that microscopic reversibility holds only in average,
that transitions at the microscopic level have unexpected properties
with important consequences for biology and life~\footnote{Ideas such as
{\em purposiveness} of changes have appeared recurrently in the context
of natural selection in biology, see for instance~\cite{MarSag95}.}. If
this were the case, while the average behavior would be well described
by current dynamical theories, rare fluctuations might display a more
refined pattern beyond our current expectation.

Biological matter tends to organize reaching fantastic levels of
complexity. Although it is often tacitly assumed that our current
understanding of physics will provide clues to fill into the many
``details'' that surround the organization of biological matter, the
truth is that bold ideas will be probably needed to go beyond the
present state of the art.  Biological matter will become a common
laboratory for physicists in order to test and understand many of the
questions that transcend the behavior of ordinary matter. Single molecule
experiments have opened a vein of research for physicists, that require
the combination of a general knowledge of physics, chemistry and biology
to grasp the most relevant aspects required to unravel the behavior of
living matter at the most fundamental level.

\noindent{\bf Acknowledgments.} I acknowledge the warm hospitality of the
Bustamante, Tinoco and Liphardt labs at UC Berkeley where this work has
been done. I thank C. Bustamante, D. Collin, C. Jarzynski, S. Smith,
I. Tinoco and E. Trepagnier for useful discussions. I wish to thank also
J. Liphardt for discussions and a critical reading of the
manuscript. This work is supported by the David and Lucile Packard
Foundation, the European community (STIPCO network), the Spanish research
council (Grant BFM2001-3525) and the Catalan government.

\vspace{.3cm}
\noindent
{F\'elix Ritort\\   
Departament of Physics\\
Faculty of Physics\\ 
University of Barcelona\\ 
Diagonal 647\\ 
08028 Barcelona, Spain \\
{\it and}\\
Department of Physics\\ 
University of California\\ 
Berkeley CA 94720, USA\\
email~:~ritort@ffn.ub.es}

\newpage

\end{document}